\documentclass[useAMS,usenatbib]{mn2e}
\usepackage{journals}
\usepackage{graphicx}
\usepackage{natbib}

\title[Optical spectroscopy of the quiescent counterpart to
  EXO\,0748$-$676]{Optical spectroscopy of the quiescent counterpart to EXO\,0748$-$676}

\author[Bassa et al.]  {C.\,G.\,Bassa$^{1,2}$\thanks{email:
    c.bassa@sron.nl}, P.\,G.\,Jonker$^{1,3}$, D.\,Steeghs$^{4,3}$, M.\,A.\,P.\,Torres$^3$\\
$^1$SRON, Netherlands Institute for Space Research, Sorbonnelaan 2, 3584\,CA, Utrecht, The Netherlands\\
$^2$Department of Astrophysics, IMAPP, Radboud University Nijmegen, Toernooiveld 1, 6525\,ED, Nijmegen, The Netherlands\\
$^3$Harvard-Smithsonian Center for Astrophysics, 60\,Garden St., Cambridge, MA\,02138, USA\\
$^4$Department of Physics, University of Warwick, Coventry CV4\,7AL, UK\\
}

\begin{document}

\date{Accepted 1988 December 15. Received 1988 December 14; in original form 1988 October 11}

\pagerange{\pageref{firstpage}--\pageref{lastpage}} \pubyear{2002}

\maketitle

\label{firstpage}

\begin{abstract} We present phase resolved optical spectroscopy and
  X-ray timing of the neutron star X-ray binary EXO\,0748$-$676 after
  the source returned to quiescence in the fall of 2008. The X-ray
  light curve displays eclipses consistent in orbital period, orbital
  phase and duration with the predictions and measurements before the
  return to quiescence. H$\alpha$ and He\,I emission lines are present
  in the optical spectra and show the signature of the orbit of the
  binary companion, placing a lower limit on the radial velocity
  semi-amplitude of $K_2>405$\,km\,s$^{-1}$. Both the flux in the
  continuum and the emission lines show orbital modulations,
  indicating that we observe the hemisphere of the binary companion
  that is being irradiated by the neutron star. Effects due to this
  irradiation preclude a direct measurement of the radial velocity
  semi-amplitude of the binary companion; in fact no stellar
  absorption lines are seen in the spectrum. Nevertheless, our
  observations place a stringent lower limit on the neutron star mass
  of $M_1>1.27$\,M$_\odot$. For the canonical neutron star mass of
  $M_1=1.4$\,M$_\odot$, the mass ratio is constrained to
  $0.075<q<0.105$. \end{abstract}


\section{Introduction}
Neutron star low-mass X-ray binaries (LMXBs) consist of a neutron star
that accretes matter from a low-mass companion star. From binary
evolution models, we know that the neutron star could accrete up to
0.7\,M$_\odot$ from the companion (e.g.\ \citealt{hb95}). In that
case, neutron stars as massive as $\ga2$\,M$_\odot$ may be expected in
LMXBs. Such a high-mass neutron star would pose strong constraints on
the equation of state (EOS) of matter at the pressures and densities
encountered in neutron star cores (see \citealt{lp04} and references
therein). Determining the neutron star EOS is one of the ultimate
goals of the study of neutron stars.

In order to determine the neutron star mass in these effectively
single-line spectroscopic binaries, we need to know the inclination,
the projected rotational velocity $v\sin i$ of the companion star and
the semi-amplitude $K_2$ of its radial velocity curve.  In principle,
these last two parameters can be determined via optical
spectroscopy. Unfortunately, the companion stars of LMXBs are nearly
always intrinsically faint and they are typically located at distances
of several kilo-parsecs. Furthermore, in systems accreting at high
accretion rates the accretion disc out-shines the small companion
stars making it impossible to detect the stellar absorption lines
necessary to determine $K_2$ and $v\sin i$. In LMXBs the binary
inclination can be determined accurately if the system is eclipsing
\citep{hor85}. Hence, in order to provide the strongest, model
independent constraints on the neutron star mass we need to target
quiescent, eclipsing LMXBs.

EXO\,0748$-$676 is such an eclipsing LMXB. It was discovered using
observations of the {\it European X-Ray Observatory Satellite}
(EXOSAT) in 1985 \citep{pwgh85}. The presence of X-ray eclipses and
type\,I X-ray bursts marking the compact object as a neutron star,
were readily found \citep{pwgg86,ghpw86}. In hindsight the source had
been observed serendipitously by the {\it EINSTEIN} satellite as well
as during EXOSAT slews (\citealt{gc99} and \citealt{rph+99},
respectively).

The optical counterpart to the X-ray source has been found by
\citet{pwgh85}. \citet{wqhm85} report on the discovery of the optical
counterpart as well as on searches of archival optical plates showing
that the optical source was not present down to roughly 23rd magnitude
in the SRC J plate when the X-ray source was inactive. \citet{cpm02}
reported on a measurement of the gravitational redshift from X-ray
absorption line spectroscopy during type~I X-ray bursts providing
$M/R$; this is, so far, unique to this source. However, these features
were not detected in subsequent observations and thus require further
confirmation (see \citealt{cpm+08}). Recently, \citet{gcl09} announced
the discovery of nearly periodic oscillations during two type I X-ray
bursts in EXO 0748-676 at a frequency of 552\,Hz. This is at odds with
the 45\,Hz burst oscillations reported by \citet{vs04}. If 552\,Hz is
indeed the neutron star spin frequency the redshift measurement is
challenged further since rotational Doppler broadening tends to lower
the contrast unless the orientation of the neutron star is favorable,
with the spin axis pointing in our direction. Using a combination of
observations, including the contested redshift measurement, and
theoretical findings \citet{oze06} argues that the mass in
EXO\,0748$-$676 is $2.10\pm0.28$\,M$_\odot$, which if confirmed would
rule out many soft equations of state. However, a high spin frequency,
such as 552\,Hz, besides challenging the redshift measurement, also
introduces sizeable relativistic corrections to the equations and thus
result of \citet{oze06}. 

Attempts to use narrow He, C and N emission lines that are probably
originating on the heated inner hemisphere of the companion star to
get a lower limit on the mass of the neutron star are presented in
\citet{phs+06} and \citet{mco+09}. The X-ray eclipse duration for
EXO\,0748$-$676 in outburst has been measured from X-ray studies using
the EXOSAT and XMM-{\it Newton} satellite (8.3 minutes,
\citealt{pwgg86} and \citealt{hwb03}, respectively).

Recently, the system returned to quiescence after a more than 20
year-long active phase (see Wolff et
al.\,2008ab\nocite{wrw08b,wrww08}, \citealt{hj08,tjss08}). Analysing
the optical $R$-band and near-infrared $J$-band light curves,
\citet{hj09} find that the residual X-ray luminosity of the cooling
hot neutron star (see \citealt{dww+09}) is causing the inner
hemisphere of the companion star to be heated.

We here present phase-resolved optical spectroscopy and X-ray
observations of the source shortly after the end of the outburst.

\section{Observations and data reduction}
\subsection{X-ray observations}
We have observed EXO\,0748$-$676 on Nov.~6, 2008 at 08:53:10 (UTC)
with the European Photon Imaging Camera (EPIC) on board the
XMM-\textit{Newton} satellite. These data were acquired 62 days after
the first X-ray observations of EXO\,0748$-$676 by
\citet{dww+09}. Hence, the source was in quiescence. The observation
time was awarded to our Directors Discretionary Time proposal. The
observation identification number is 0560180701. Two of the three
imaging cameras on board XMM-\textit{Newton} use Metal Oxide
Semi-conductor (MOS) CCD arrays whereas one uses a pn CCD array. Since
the MOS cameras receive less than 50 per cent of the light due to the
presence of the Reflection Grating Spectrometers in the light path we
present results of the pn camera only. The pn camera was employed in
\textsl{PrimeFullWindow} mode and a medium filter was put in place to
avoid strong contamination due to bright optical sources in the field
of view.

To benefit from the latest calibrations available in December 2008 we
have reprocessed and analysed the data using the \textit{SAS} software
version 8.0.0.  Data are excluded if the 10-12\,keV background count
rate for single events in a $3\times3$ CCD-pixel region (with pattern
0.0) is higher than 1.0\,count\,s$^{-1}$. The net on-source exposure
time is 29.82\,ks. We have produced bary-centered events lists using
the \textit{SAS} tool \textsl{barycen}.

Next, we have produced a background subtracted source light curve with
10\,second time bins including events of photon energies larger than
0.15\,keV. Source photons have been extracted from a circular region
of $35\arcsec$ radius centered on the best known source
position. Background photons have been extracted from a source-free
region of the same size and shape on the same CCD.

\begin{figure}
  \includegraphics[width=8cm]{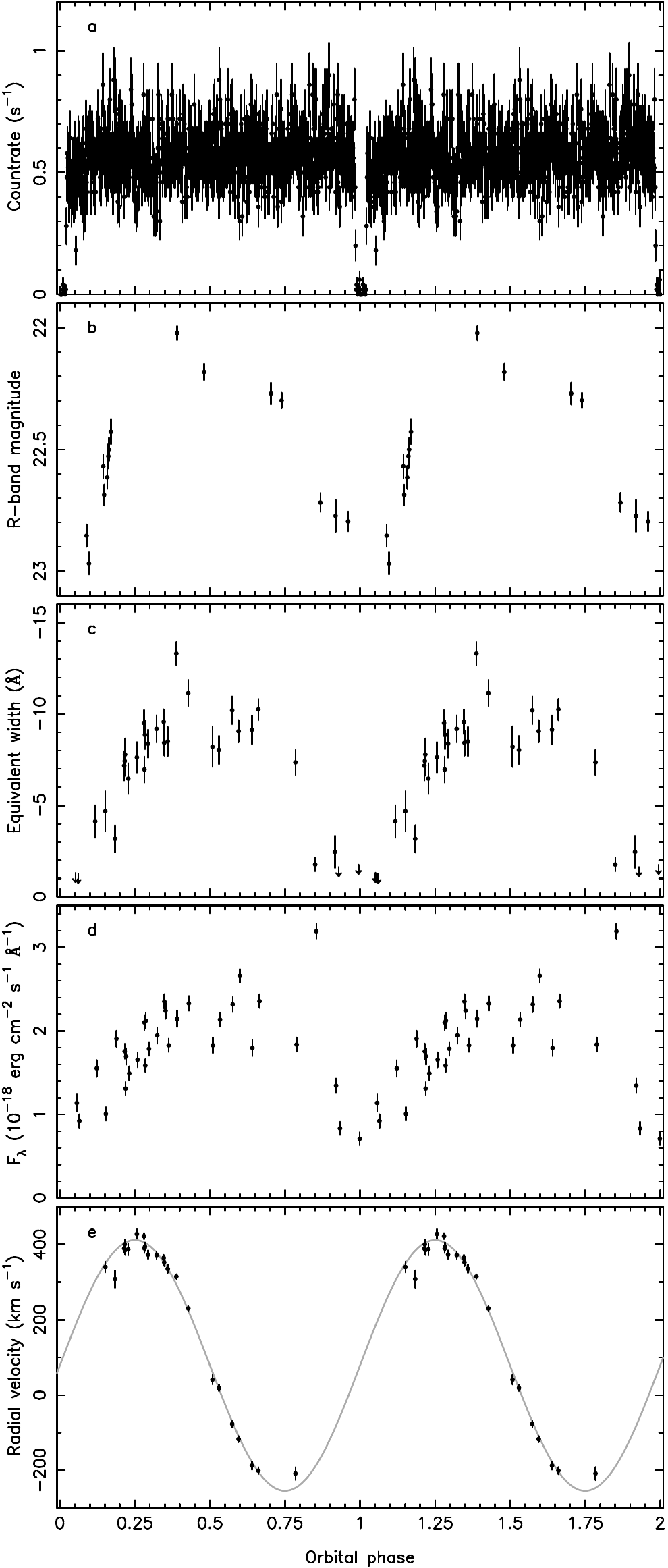}
  \caption{X-ray and optical variability and the radial velocity
    folded on the constant-period orbital ephemeris of
    \citet{whw+02}. Two orbits are shown for clarity. \textit{(a:)} The
    X-ray light curve showing the eclipse at phase
    $\phi=0$. \textit{(b:)} $R$-band magnitudes determined from the
    acquisition images. \textit{(c:)} Equivalent width of the narrow
    H$\alpha$ component, determined from Gaussian fits (see
    \S\,\ref{ss:radvel}). \textit{(d:)} The average optical flux from
    the individual spectra between 6000\,\AA\ and 6250\,\AA. Both in
    broadband filters and narrow lines, the optical flux of the
    optical counterpart varies with the orbital period, being faintest
    at the time of the X-ray eclipse. Note that variable slit losses
    are likely present. \textit{(e:)} Radial velocities
    determined by fitting Gaussians to the H$\alpha$ profile. Only
    measurements between $0.15<\phi<0.85$ are shown. The grey curve
    shows the best fit circular orbit
    ($v(\phi)=\gamma+K_\mathrm{em}\sin 2\pi \phi$) to the data.}
  \label{fig:var}
\end{figure}

\subsection{Optical observations}
Long-slit spectra of the optical counterpart to EXO\,0748$-$676 were
obtained with FORS2, the low dispersion spectrograph of ESO's Very
Large Telescope. Between November 25th and December 26th, 2008, a
total of 31 spectra were obtained, all with 870\,s exposure times and
using the 1200R grism, covering the wavelength range between
5750\,\AA\ and 7310\,\AA, and a $0\farcs7$ slit. The detectors were
read out with $2\times2$ binning, providing a resolution of
2.1\,\AA\ sampled at a dispersion of 0.76\,\AA\,pix$^{-1}$.  All
spectra were obtained during times of good to excellent seeing
($0\farcs48$ to $0\farcs75$).

The images were corrected for bias and flat-fielded using lamp
flats. For the sky subtraction we used clean regions on the slit,
fitting it with a polynomial of up to 2nd order if that provided the
best fit. The spectra were optimally extracted following the algorithm
of \citet{hor86} and wavelength calibrated using arc lamp exposures
taken during daytime. Finally, an approximate flux calibration, not
correcting for slit losses, was determined from a 22\,s exposure of
the spectrophotometric standard LTT\,3218 taken on the first night
with a $5\arcsec$ slit.

We furthermore analysed 15 $R$-band acquisition images of 30 or 40\,s
exposure times taken prior to the spectral observations. These images
were corrected for bias and flat-fielded using sky flats. Instrumental
magnitudes of stars on each image were determined through
point-spread-function fitting using DAOphot\,II \citep{ste87}. All
magnitudes were calibrated against the instrumental magnitudes of
stars on the first image by using some 350 to 400 comparison stars to
fit for an offset in instrumental magnitude.

For the photometric calibration of the first image, we used aperture
photometry of a few bright stars to determine the aperture correction
and determined the photometric calibration from 19 photometric
standards from the PG\,2213$-$006 field \citep{lan92,ste00}. We used
the standard FORS2 $R$-band extinction coefficient of 0.08\,mag per
airmass to correct for the difference in airmass between the field of
EXO\,0748$-$676 and the PG\,2213$-$006 standard field. Because no
color coefficients could be determined, we estimate the uncertainty in
the photometry as the rms in the zeropoint, which is about 0.1\,mag.

\section{Data analysis}

\subsection{X-ray timing}\label{ss:timing}
During the X-ray observations of EXO\,0748$-$676 the source shows a
constant count rate of $0.55\pm0.01$\,counts\,s$^{-1}$ except for the
two distinct eclipses. Fig.\,\ref{fig:var}a shows the X-ray
light-curve of the total detected counts (including background). The
background flux was consistent with being constant at
0.015\,counts\,s$^{-1}$. For the present work we are mainly interested
in the eclipse duration and the orbital period and mid eclipse
times. The spectral properties of the source will be presented
elsewhere. We fit the light curve by a constant plus four step
functions located at $T_0-\frac{1}{2}\Delta t$ and
$T_0+\frac{1}{2}\Delta t$ for the first eclipse and
$T_0+P_\mathrm{b}-\frac{1}{2}\Delta t$ and
$T_0+P_\mathrm{b}+\frac{1}{2}\Delta t$ for the second eclipse. Here
$T_0$ is the mid eclipse time of the first eclipse, $P_\mathrm{b}$ is
the orbital period and $\Delta t$ is the eclipse duration. The step
functions are approximated by heaviside functions, defined as
$f(t)=\frac{1}{2}+\frac{1}{2}\tanh(kt)$, where $k=-400$\,hour$^{-1}$
is used to describe the eclipse ingress and $k=400$\,hour$^{-1}$
describing the eclipse egress. With these values for $k$,
50\,per\,cent of the flux at the ingress and egress times is contained
in a single 10\,s bin.  Fitting for $T_0$, $P_\mathrm{b}$, $\Delta t$
and a constant to describe the flux out of the eclipse, we obtain
$T_0=54776.501663\pm0.000068$ MJD/TDB,
$P_\mathrm{b}=3.8276\pm0.0018$\,h and $\Delta t=500\pm7$\,s.

We have compared $T_0$ and $P_\mathrm{b}$ with the values predicted by
the orbital ephemerids of EXO\,0748$-$676 obtained from RXTE X-ray
timing by \citet{whw+02}. Extrapolating their ephemerids to orbit
$n=54384$ gives consistent results for the orbital period at the
$2\sigma$ level, while the mid eclipse time falls between the
predictions of the constant-period ephemeris and the quadratic
ephemeris using all data. At $6\sigma$, the prediction from the
constant-period ephemeris lies closest to our measurement. Because the
ephemeris of \citet{whw+02} uses a considerably larger dataset, their
results are expected to represent the long term timing better, and
hence, for the remainder of the paper we will use the mid eclipse time
$T_0$ and orbital period $P_\mathrm{b}$ given by the constant-period
ephemeris of \citet{whw+02}.

We selected in total 800\,s worth of data centered on the mid-eclipse
times for both eclipses in our dataset. Even during the X-ray
eclipses, EXO\,0748$-$676 is still detectable, providing 29 source
counts (38 total counts from which 9 are expected due to the
background) in the 800\,s of observation that we used. This
constitutes approximately 6.6\,per\,cent of the out of eclipse
$0.3-10$\,keV count rate.

\subsection{Radial velocities and spectral properties}\label{ss:radvel}
\begin{figure*}
  \includegraphics[width=8cm]{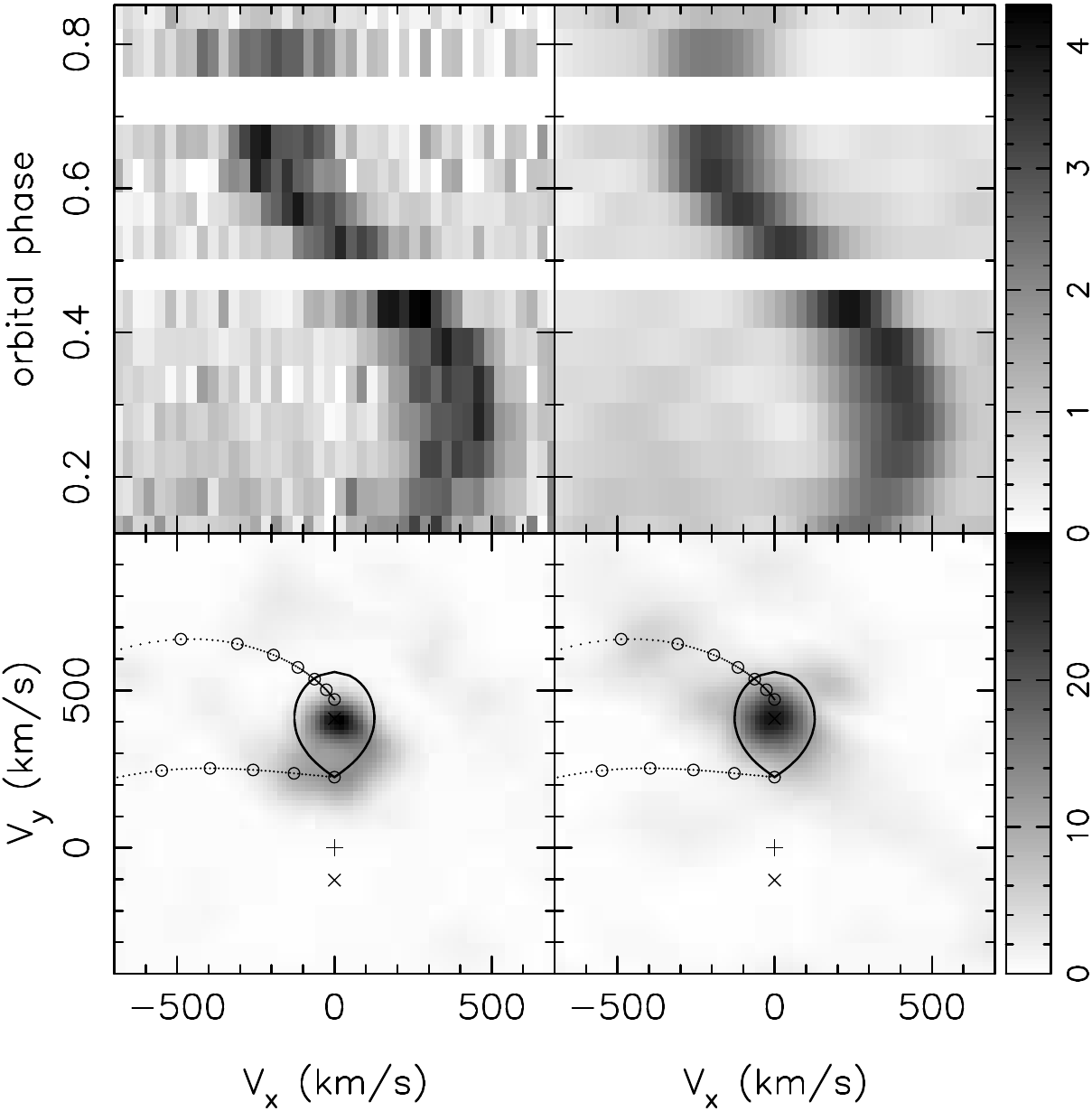}
  \includegraphics[width=8cm]{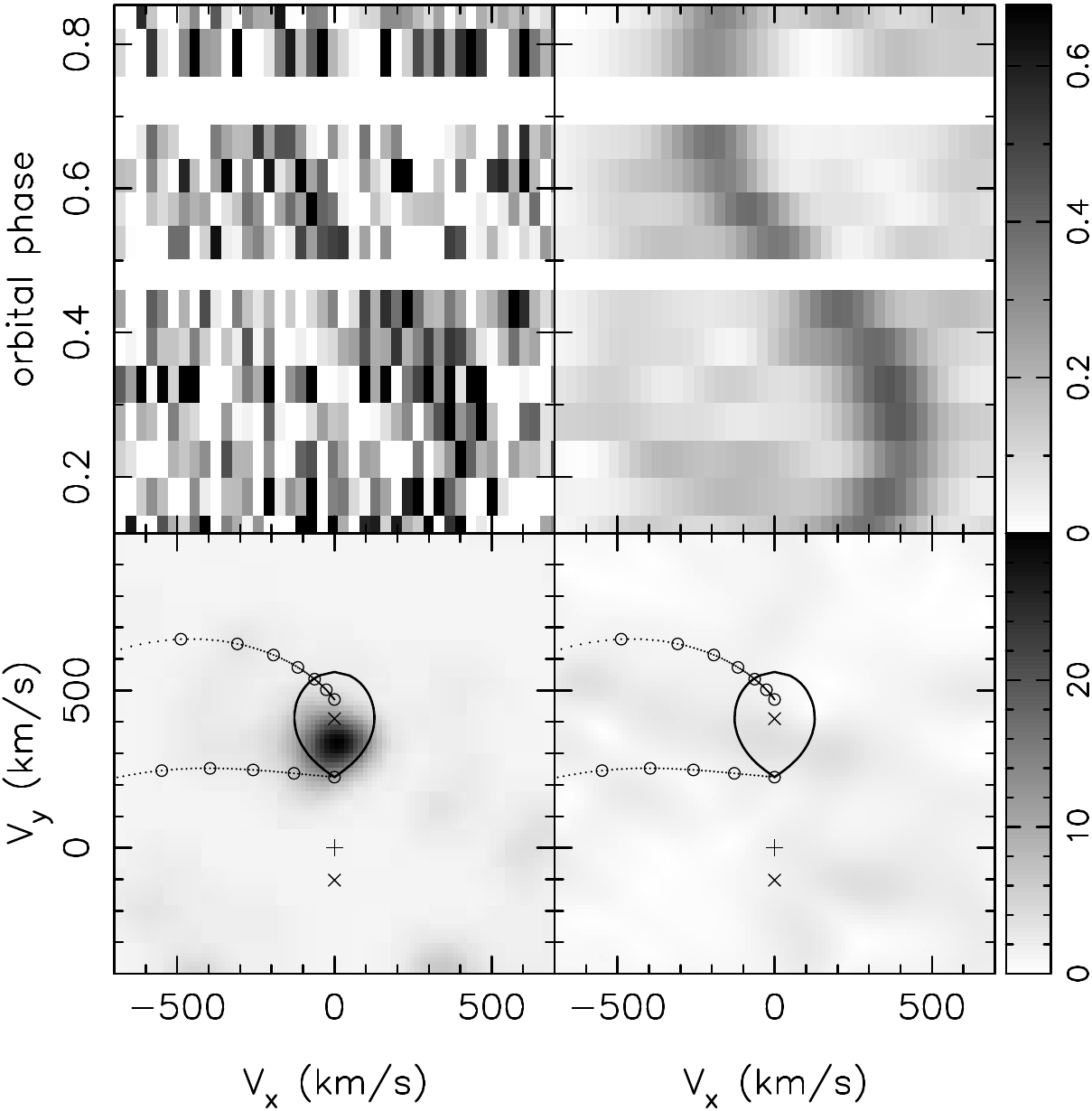}
  \caption{Tomographic reconstructions for H$\alpha$ \textit{(left
      panel)} and He\,I $\lambda6678$ \textit{(right panel)}. In each
    case, the top-left panel shows the observed line profiles as a
    function of binary phase. The top-right panel shows the
    reconstructed data from the converged maximum-entropy
    solution. Both are displayed on the same gray scale as shown in
    the wedge. Bottom panels show the corresponding tomograms
    resolving the line emission across the $V_x-V_y$ plane. Left is
    the constant contribution to the flux, right the phase-dependent
    component (see \citealt{ste03}). The gray scale wedge denotes the
    fractional amplitude of the variable map in per cent. For
    reference a Roche lobe model and ballistic stream trajectory is
    plotted for an assumed system mass ratio of $q=0.25$ and projected
    donor star orbital velocity of $K_2=410$\,km\,s$^{-1}$.}
  \label{fig:doptom}
\end{figure*}

The individual extracted optical spectra are generally featureless
except for the region around H$\alpha$ where all spectra have a broad
emission component (FWHM of about 26\,\AA), while a strong narrow
emission line (5\,\AA\ FWHM) is present in spectra taken between
orbital phases of approximately $0.15<\phi<0.85$. Here $\phi=0$ is
defined as the center of the X-ray eclipse. Fig.\,\ref{fig:doptom}
shows a trailed spectrum of the H$\alpha$ region, folded on the
ephemeris of \citet{whw+02}, where the emission clearly follows the
orbital signature of a circular orbit. The emission appears strongest
between the quadratures and weakest at the time of the X-ray eclipse.

To measure radial velocities of the H$\alpha$ components we fit the
spectral range between 6400\,\AA\ and 6700\,\AA\ of each spectrum with
either a single Gaussian for spectra taken between $\phi>0.85$ and
$\phi<0.15$ and two Gaussians for spectra taken between
$0.15<\phi<0.85$. Three parameters, position, width and height, were
fitted for each Gaussian. Fitting a circular orbit with
$v(\phi)=\gamma+K_\mathrm{em}\sin 2\pi\phi$ to the radial velocities
of the narrow emission component returns
$\gamma=78.6\pm3.9$\,km\,s$^{-1}$ and
$K_\mathrm{em}=333.0\pm5.0$\,km\,s$^{-1}$ with a reduced
$\chi^2_\mathrm{r}=2.4$ for 22 degrees of freedom (see
Fig.\,\ref{fig:var}e). The errors on the values given above are the
formal $1\sigma$ uncertainties and are scaled to give
$\chi^2_\mathrm{r}=1$. Because these radial velocities are determined
from emission lines the measured radial velocity semi-amplitude
$K_\mathrm{em}$ may or may not be equal to the true radial velocity
semi-amplitude of the companion $K_2$, see \S\,\ref{s:discon}.

For this fit, the orbital period $P_\mathrm{b}$ and mid eclipse time
$T_0$ were set to those of the ephemeris of \citet{whw+02}. Leaving
these parameters free in the fit gives values that are fully
consistent with their constant-period ephemeris extrapolated to
$n=54595$. The difference in $T_0$ is only $1.1\pm0.7$\,min
($0.005\pm0.003$ in phase), while the difference in $P_\mathrm{b}$ is
$0.18\pm0.38$\,s.

\begin{figure*}
  \includegraphics[width=17cm]{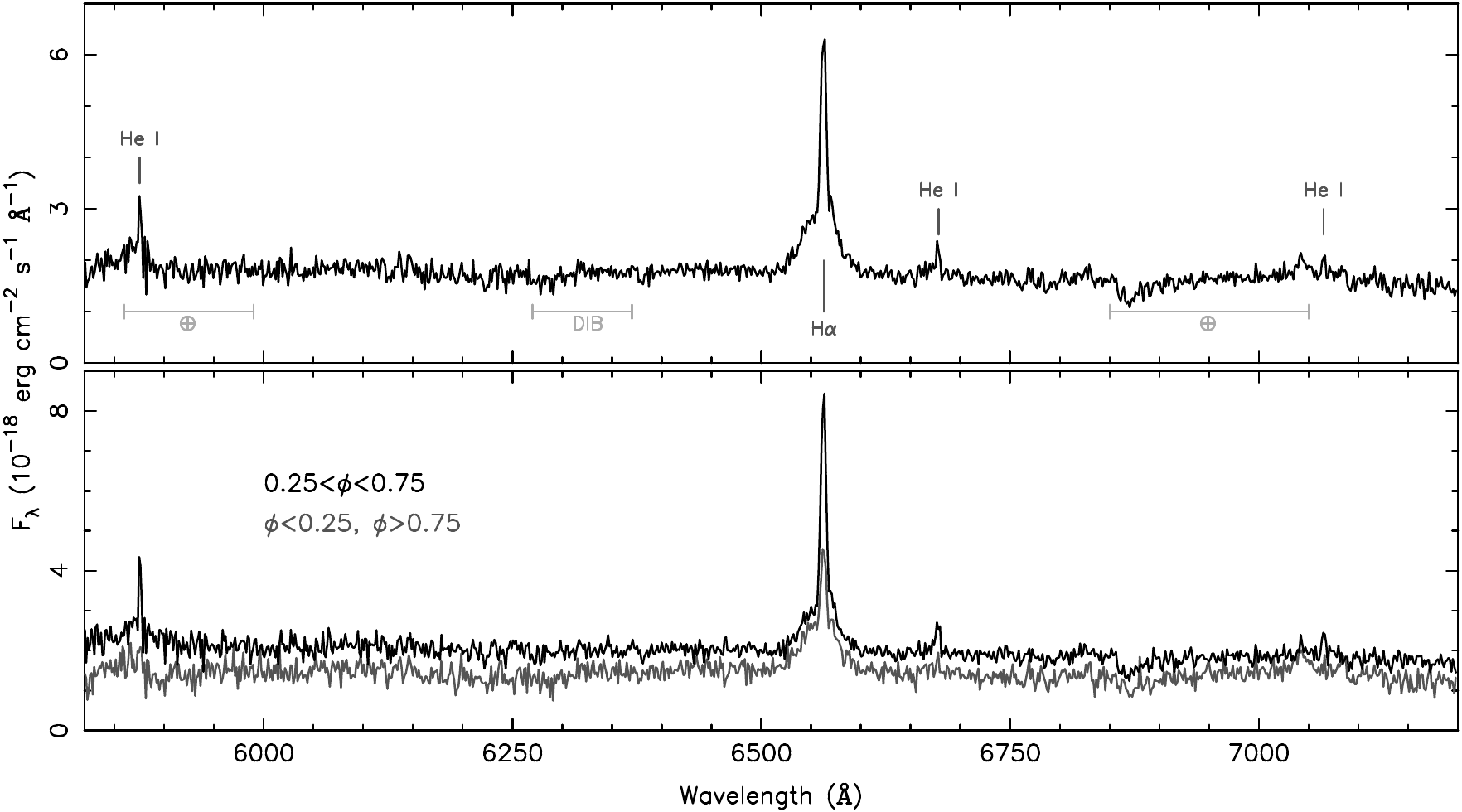}
  \caption{\textit{(top:)} The spectrum of the counterpart to
    EXO\,0748$-$676. This spectrum is the average of the 31 individual
    spectra, each shifted to zero velocity using the orbital fit to
    the narrow H$\alpha$ emission. The broad and narrow H$\alpha$
    components, as well as three He\,I emission lines are
    visible. Regions of telluric absorption and diffuse interstellar
    bands are indicated. \textit{(bottom:)} Zero velocity averages of
    the individual spectra with orbital phases between
    $0.25<\phi<0.75$, basically showing the spectrum from the
    hemisphere of the companion that is facing the neutron star, and
    between $\phi>0.75$ and $\phi<0.25$, showing the spectrum of the
    opposite hemisphere. It shows that the emission lines are
    strongest on the hemisphere facing the neutron star, and that that
    hemisphere is brighter by about a factor 2.}
  \label{fig:spec}
\end{figure*}

An average of the 31 individual spectra is shown in
Fig.\,\ref{fig:spec}, where the individual spectra are shifted to zero
radial velocity using the H$\alpha$ radial velocity curve. This
spectrum shows, besides the narrow and broad H$\alpha$ components,
weak emission lines of He\,I $\lambda5875$, $\lambda6678$ and
$\lambda7065$. Some telluric absorption features are present at
$6800$\,\AA. Zero velocity averages of individual spectra with orbital
phases between $0.25<\phi<0.75$ and between $\phi>0.75$ and
$\phi<0.25$ are also shown in Fig.\,\ref{fig:spec}. Here, the first
spectrum predominantly shows the hemisphere of the companion facing
the neutron star, while the latter spectrum shows the opposite
hemisphere. The spectra are significantly different, with the
H$\alpha$ and He\,I emission lines being strongest on the hemisphere
facing the neutron star. Furthermore, that hemisphere appears brighter
by about a factor 2.

Though no absorption features appear to be present that may be
attributable to the binary companion, we have compared the averaged
spectrum against synthetic stellar spectra by \citet{mscz05}. In the
wavelength range covered by our observations, the strongest absorption
features are the TiO bands around 6180\,\AA\ and 7050\,\AA\ for
temperatures below about $T_\mathrm{eff}=4000$\,K, while the Na\,I\,D
lines are present for temperatures between 4000\,K to 7500\,K,
decreasing in strength with increasing temperature. Between 4000\,K to
5000\,K many metallic lines are present in the range between
6000\,\AA\ to 6400\,\AA. Above about 5000\,K H$\alpha$ becomes
increasingly stronger.

Unfortunately the region around the Na\,I\,D lines overlaps with the
He\,I $\lambda5875$ emission line while the H$\alpha$ emission likely
swamps the absorption component. As such, both regions cannot be
used. Instead, we use the region between 6000\,\AA\ and 6250\,\AA,
computing a $\chi^2$ statistic by fitting the ratio in flux between
the template and the zero velocity averaged spectrum with a second
order polynomial. We find that the best fit solution sets a $3\sigma$
limit on the temperature of $T_\mathrm{eff}\ga5000$\,K, regardless of
the surface gravity. This temperature is the temperature above which
the TiO bands, disappear. The fit has $\chi^2_\mathrm{r}=1.5$ for some
4900 degrees of freedom. The absence of absorption lines in the
observed and synthetic spectra essentially prevent us from obtaining a
better constraint on the temperature of the companion, and measuring
its radial velocity curve using photospheric absorption lines.

\subsection{Variability}\label{ss:var}
The spectroscopic as well as the photometric observations show that
the optical flux varies in both the continuum and emission
lines. Fig.\,\ref{fig:var}b shows the $R$-band magnitudes determined
from the acquisition images and they confirm the light curve published
by \citet{hj09}; a single peaked modulation with the minimum at the
time of the X-ray eclipse. This trend is also present in both the
equivalent width of H$\alpha$ (Fig.\,\ref{fig:var}c) and the continuum
emission (Fig.\,\ref{fig:var}d). The narrow H$\alpha$ emission seems
to disappear completely near the phase of the X-ray eclipse at
$\phi=0$, while the broadband flux varies by a factor of about 2.

\subsection{Doppler tomography}\label{ss:tomo}
In order to prepare the spectra for emission line tomography, we first
continuum normalised each spectrum using a spline-fit and re-binned
the data to a constant velocity scale sampling of
33.8\,km\,s$^{-1}$\,pix$^{-1}$. The spectra were then phase-binned
using the \citet{whw+02} ephemeris. The prominent orbital variation in
the H$\alpha$ component led us to employ the modulation Doppler
tomography method described in \citet{ste03}. This technique includes
variable contributions modulated on the orbital period and was used to
construct emission line tomograms resolving the H$\alpha$ and He\,I
$\lambda6678$ emission distributions. We used a systemic velocity of
78\,km\,s$^{-1}$ as derived in \S\,\ref{ss:radvel}, although the
reconstructions are robust against the exact choice of $\gamma$. The
advantages of Doppler tomography are its ability to map complex line
profiles using the whole data set at once. Contributions from specific
sources in the binary can then be identified by their known position
on the tomogram.

Reconstructions achieved good fits to the observed line profiles,
achieving formal $\chi_\mathrm{r}^2=1.1$ for both H$\alpha$ and He\,I
$\lambda6678$. We show the reconstructions for these lines in
Fig.~\ref{fig:doptom}. The bulk of the emission originated from a
region consistent with the location of the mass donor star, although a
weak underlying signature from an accretion disc can also be
seen. These are responsible for the broad base (disc) and narrow
(donor) moving line components previously mentioned. Since we are
mainly concerned with the contribution from the donor, we restrict the
plotted velocity intervals to the region surrounding the expected
location of the donor.

In H$\alpha$, the donor star component is reconstructed as an extended
feature centred on the Y-axis, covering velocities between 200 and
410\,km\,s$^{-1}$. This emission is furthermore significantly
modulated with an amplitude of 30\,per\,cent, peaking near orbital
phase $\phi=0.5$. Both the distribution as well as the modulation of
this emission is exactly what would be expected for line emission from
an irradiated donor star. However, the emission does not appear to be
uniformly distributed across the Roche lobe. We measured the centroid
position of the pronounced peak in our Doppler tomograms using several
methods and found consistent velocity measurements to within a few
km\,s$^{-1}$. For example, a Gaussian fit to the H$\alpha$ map
delivers a peak of $V_y=K_\mathrm{em}=413$\,km\,s$^{-1}$ while the
highest pixel is at 410\,km\,s$^{-1}$. We also reconstructed Doppler
maps using different values for the systemic velocity to see by how
much the measured peak position would be affected. Varying the
systematic velocity $\gamma$ by as much as 15\,km\,s$^{-1}$ shifts the
peak velocity down by up to 5\,km\,s$^{-1}$. The donor star component
is broadened and weaker for reconstructions with $\gamma$
significantly above or below our assumed value of
78\,km\,s$^{-1}$. This is further confirmation that this is close to
the true value for the systemic velocity and that its impact on the
measured $K_\mathrm{em}$ is small. Given that formal error propagation
is not possible in maximum-entropy regularised inversions such as our
Doppler maps, we conservatively assign an uncertainty of
5\,km\,s$^{-1}$ to our measured velocities. This uncertainty
encompasses the observed scatter in derived $K_\mathrm{em}$ when
comparing different centroid methods as well as the systematic
contribution due to a possible uncertainty in $\gamma$ (as evaluated
above). 

For H$\alpha$, our measured $K_\mathrm{em}=410\pm5$\,km\,s$^{-1}$ is
significantly larger than the $K_\mathrm{em}$ values determined in
\S\,\ref{ss:radvel} via our radial velocity curve fits. For
comparison, He\,I $\lambda6678$ shows a more spot-like distribution,
again centred closely to the Y-axis and peaking at
$V_y=K_\mathrm{em}=345\pm5$\,km\,s$^{-1}$. We do not have sufficient
signal-to-noise in this much weaker line to constrain its variability
amplitude to a significant degree. We thus find that the peak of the
H$\alpha$ emission is found further back on the Roche lobe compared to
the He\,I emission. For illustration, we plot the expected shape of
the donor star Roche lobe in Fig.~\ref{fig:doptom} using a mass ratio
of $q=0.25$ and $K_2=410$\,km\,s$^{-1}$.  The various estimates for
$K_\mathrm{em}$ will be compared and discussed in the next section.

\section{Discussion and conclusions}\label{s:discon} We have obtained X-ray
timing and phase-resolved optical spectra of the eclipsing low-mass
X-ray binary EXO\,0748$-$676 shortly after the source transitioned
into quiescence after several decades of outburst activity. Timing of
the X-ray eclipses show that the orbital period and eclipse duration
are consistent with the values when the source was in outburst. In
addition, the orbital phase is consistent with extrapolating the
constant period ephemeris of \citet{whw+02}.

The X-ray eclipses in EXO\,0748$-$676 are near total, however, about 6
to 7 per cent of light remains during the eclipses. \citet{bhf+01}
find that in outburst part of this residual flux can be attributed to
fluorescent thermal halo emission. \citet{hwb03} find that in outburst
about 4\,per\,cent of the $2-6$\,keV continuum emission is detected
during eclipse. They attribute this to light scattered into our line
of sight by a corona. Given that they find that the fraction of light
during the eclipse is larger for lower energies, our value seems
consistent with that found in outburst. This implies that the
structure responsible for the scattering is still present, or still
replenished even when the source is in quiescence.

Using our medium resolution VLT/FORS2 spectra of EXO\,0748$-$676 we
find emission lines of H$\alpha$ and He\,I. They display the orbital
motion expected if the lines originate on the binary companion. The
optical spectra show considerable flux variations in both the
continuum and the emission lines. The variations display a single peak
over the orbital period and we attribute this to irradiation of the
hemisphere of the companion facing the neutron star. \citet{hj09} also
presented evidence for an irradiated donor star on the basis of their
optical and near-infrared photometric observations.

This irradiation causes three effects that will skew the measurement
of the radial velocity semi-amplitude. First, due to the irradiation
the centre of light is offset from the centre of mass of the
companion. The former being on the hemisphere facing the neutron star,
the measured radial velocity semi-amplitudes are smaller than the
actual orbital velocity of the companion. To correct for this
difference one can apply a so called '$K$-correction'
\citep{wh88}. This '$K$-correction' can be estimated assuming values
for the mass-ratio $q$ and the opening angle of the accretion
disk. The latter is needed to correct for the partial shielding of the
binary companion by the accretion disc from the irradiating flux
(cf.~\citealt{mcm05}). Second, the irradiation may induce gas motion
on the surface of the mass donor star, resulting in apparent shifts in
orbital phase of the radial velocity curve (\citealt{kir82};
cf.~\citealt{jsnk05}). Third, due to the difference in flux observed
from the irradiated and non-irradiated hemispheres of the companion
the rotational broadening profiles are not symmetric. As a result
observed line profiles can become dependent on orbital phase and
inclination (\citealt{jkg03}; \citealt{jsnk05}). This effect is
obvious when considering the extreme case where all the flux
originates from the irradiated hemisphere. Assuming synchronous
rotation, at phase $\phi=0.25$ the companion is moving away from the
observer at its maximal observed orbital velocity, however, the
observed irradiated hemisphere is rotating towards the observer
lowering the observed semi-amplitude of the radial velocity curve.

How much the measurement of the radial velocity semi-amplitude is
influenced depends therefore on whether the semi-amplitude is
determined by fitting Gaussians to the emission line profiles or by
fitting a two dimensional Gaussian to the spot in a Doppler tomogram.
We modelled the changes in the line velocities by fitting a double
Gaussian, one narrow, one significantly broader, to the H$\alpha$
emission profiles. We find that the velocities of the narrow component
can be well described by fitting a sinusoid with a constant. The
semi-amplitude is $338.8\pm4.8$\,km\,s$^{-1}$ and the constant
representing the systemic velocity $\gamma=78.5\pm3.4$\,km\,s$^{-1}$
when restricting the fit to the velocities over the orbital phases
$0.25<\phi<0.75$ when the narrow component of the emission line is
strongest.

Using Doppler tomography of the H$\alpha$ emission line with
$\gamma=78$\,km\,s$^{-1}$ as input we find that the centre of the
H$\alpha$ emission spot in the Doppler map falls at
$K_\mathrm{em}=410\pm5$\,km\,s$^{-1}$.  This value for the radial
velocity semi-amplitude measured from Doppler tomography is
significantly larger than that obtained fitting a double Gaussian to
the emission peak.  The reason for this difference lies in the fact
that the Doppler tomography method does not require symmetric line
profiles whereas fitting Gaussians obviously does. The Doppler maps do
show that H$\alpha$ emission is complex in shape and thus that the
single Gaussian model for the narrow component is likely too simplistic
to reproduce the observed distribution.  The third effect mentioned
above renders the line profiles asymmetric, affecting the determined
velocities for the 'double Gaussian' method predominantly. The Doppler
tomography will, however, suffer less from this effect yielding
velocities closer to the true centre-of-mass velocities. Note,
however, that also the Doppler tomography velocities follow the centre
of light and hence they also require an unknown '$K$-correction'.

For the only other emission line strong enough for a Doppler map
study, the He\,I $\lambda6678$ line, we obtain
$K_\mathrm{em}=345\pm5$\,km\,s$^{-1}$. There are a number of effects
controlling the distribution of line emission across the Roche lobe,
and hence defining the location of the centre of light. Parts of the
lobe may be shielded from irradiation due to a vertically extended
disc, and the optical thickness of this will be dependent on
wavelength. The incident angle of the incoming radiation is also much
larger near the terminator compared to the region near the inner
Lagrangion point (L1). Different emission lines will thus show a
different distribution reflecting the varying local conditions of the
gas, the ionisation potential of the element involved as well as the
geometry of the incoming radiation field. This typically leads to
higher-ionisation potential lines to be formed closer to the L1 point
(e.g.\ \citealt{har99,umm06}).

Our finding that He\,I emission peaks closer to the L1 point compared
to H$\alpha$ thus agrees with this picture. It shows that the Balmer
lines can still be excited near the terminator, but at the same time
also suffer from more effective shielding near the front of the lobe
because of the increased optical depth of the accretion flow due to
neutral Hydrogen.

The measured radial velocity semi-amplitudes are higher and carry
smaller statistical errors than those found during outburst by
\citet{mco+09}. Besides a higher radial velocity semi-amplitude, we
also find a higher systemic velocity, although the values are
consistent within the $3\sigma$ errors. \citet{mco+09} observed the
system in outburst using emission lines from He\,II whereas our
observations are done in quiescence and using H$\alpha$. The
differences in X-ray flux between outburst and quiescence and the
different element, may explain the difference in measured radial
velocity semi-amplitude in our data compared to \citet{mco+09} as
explained above.

Line emission from the secondary can be expected to lie anywhere
between its L1 point and the terminator. Rather than make specific
assumptions about the irradiation geometry, we will instead use the
peak of the H$\alpha$ emission from the Doppler map as a determination
of the radial velocity semi-amplitude, since it provides the strictest
lower limit on the true radial velocity semi-amplitude of the mass
donor, i.e. $K_2>K_\mathrm{em}$. Hence, with
$K_\mathrm{em}=410\pm5$\,km\,s$^{-1}$, we find that
$K_2>405$\,km\,s$^{-1}$.

\begin{figure} \includegraphics[width=8cm]{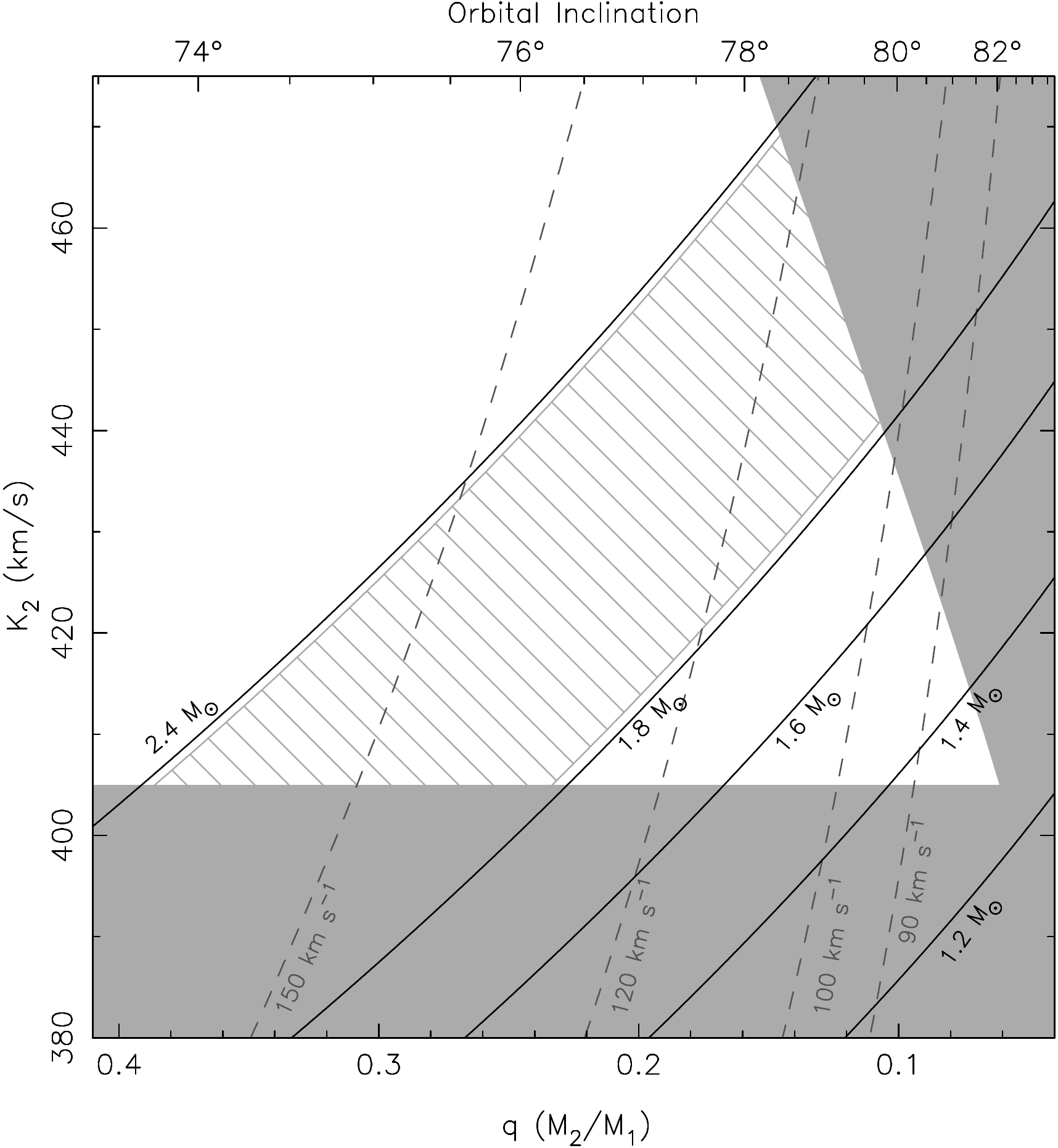}
 \caption{Constraints
    on the neutron star mass in EXO\,0748$-$676. Under the assumption
    that the companion is filling its Roche lobe, the eclipse
    measurement of the eclipse duration gives a relation between the
    mass ratio $q=M_2/M_1$ and the inclination of the orbit
    \citep{hor85}. Furthermore, the mass ratio relates to the
    rotational velocity $v\sin i$ and the radial velocity
    semi-amplitude $K_2$. The dashed lines show curves for specific
    rotational velocities. Finally, the neutron star mass function
    relates the neutron star mass to the mass ratio and the radial
    velocity semi-amplitude, giving the solid lines for specific
    neutron star masses. The horizontal grey area is excluded by our
    lower limit on the radial velocity semi-amplitude of
    $K_\mathrm{em}=410\pm5$\,km\,s$^{-1}$, setting
    $K_2>405$\,km\,s$^{-1}$, while the diagonal grey area is excluded
    in order to satisfy that all emission originates between the
    radial velocity of L1 and the center-of-mass of the companion
    ($V_\mathrm{L1}<K_\mathrm{em}<K_2$). These constraints dictate
    that the neutron star is heavier than $M_1>1.27$\,M$_\odot$. The
    hashed area is the mass range of $M_1=2.10\pm0.28$~M$_\odot$
    favoured by the work of \citet{oze06}.} \label{fig:mass}
\end{figure}

So far, we have used the fact that the true $K_2$ is larger than the
observed $K_\mathrm{em}$, where the '$K$-correction' depends on the
distribution of the emission across the Roche lobe. Thus the lines
with the largest $K_\mathrm{em}$ must be formed closest to the
terminator, whereas lines with lower $K_\mathrm{em}$ are more towards
L1. An additional limit can thus be set
by demanding that $V_\mathrm{L1}<K_\mathrm{em}$. \cite{mco+09}
reported $K_\mathrm{em}$ values that are somewhat lower than ours and
we used their lower limit on $K_\mathrm{em}=300$\,km\,s$^{-1}$ to
calculate which mass ratios are ruled out because the velocity of the
L1 point would be larger than 300\,km\,s$^{-1}$, which would be
inconsistent with the donor star components observed during
outburst. This leads to a lower limit on the system mass ratio $q$ for
a given $K_2$ which is indicated by the shaded region in the upper
right of Fig.\,\ref{fig:mass}.  From this we see that the range of
measured donor star velocities that all must satisfy
$V_\mathrm{L1}<K_\mathrm{em}<K_2$ rule out neutron star masses below
1.27\,M$_\odot$, so that $M_1>1.27$\,M$_\odot$. For the canonical
neutron star mass of $M_1=1.4$\,M$_\odot$ our observations constrain
the mass ratio to a narrow range of $0.075<q<0.105$ and the companion
mass to 0.11\,M$_\odot<M_2<0.15$\,M$_\odot$.

Narrow H$\alpha$ emission originating from the secondary star has also
been observed in other quiescent low mass X-ray binaries. When
measured after subtracting the underlying H$\alpha$ absorption line
from the secondary and accounting the veiling from the accretion disc,
the equivalent width in these systems is of 2-3 \AA\ and comparable to
that observed in rapidly rotating chromospherically active stars (see
e.g.\ \citealt{cmc+97,tcmc02,tcg+04}). In contrast, the narrow
H$\alpha$ component in EXO\,0748$-$676 reaches an equivalent width of
tenths of \AA\ (Fig.\,\ref{fig:var}). These high values of the
equivalent width can be explained as due mostly to X-ray irradiation
of the secondary.  Stellar rotation in G-M dwarfs cannot produce such
a strength because of saturation effects of stellar activity with
rotation (see e.g.\ \citealt{sshj93,sbk+97}).

Fitting theoretical stellar template spectra to the observed spectrum
averaged in the frame of the companion star, we constrain the
temperature of the inner face of the mass donor star to be larger than
5000 K. Stars of such a temperature do not show significant absorption
lines in the spectral range covered by our observations. Such
temperatures are unexpected for the orbital period of
EXO\,0748$-$676. A Roche lobe filling main sequence star would have a
mass corresponding to that of an M2 star and thus a temperature of
approximately 3500 K, much lower than the lower limit derived above.
Besides evidence from the equivalent widths of the above mentioned
emission lines and the comparison of theoretical stellar templates, we
find that the continuum varies in accord with a heated inner
hemisphere of the mass donor star (see Fig.\,\ref{fig:var}d), hence,
the variability in magnitude over the orbit cannot be explained fully
by the change in equivalent width of the emission lines.

Repeat observations targeting a different wavelength range showing
deep stellar absorption lines from the mass donor star even in the
presence of heating due to the hot neutron star have been
requested. Furthermore, the hot neutron star probably causing the
irradiating flux (\citealt{hj09}) could cool significantly on
timescales of a year reducing the irradiating flux
(\citealt{dww+09}).

\section*{Acknowledgments}
This research is based on observations collected with ESO Telescopes
at the Paranal Observatory under programme ID 282.D-5012(A) and on
observations obtained with the XMM-\textit{Newton} satellite. PGJ
acknowledges support from the Netherlands Organisation for Scientific
Research (NWO). MAPT acknowledges support from NASA grants G08-9041X
and G09-8055X. DS acknowledges a STFC Advanced Fellowship.

\label{lastpage}

\end{document}